\documentclass[twocolumn,prb,aps,nobalancelastpage,amssymb,citeautoscript,floatfix]{revtex4}

\usepackage{graphicx}
\begin{document}

\title{de Haas-van Alphen effect investigation of the \\ electronic structure of Al substituted MgB$_2$}

\author{A. Carrington,$^a$ J.D. Fletcher,$^a$ J.R. Cooper,$^b$ O.J. Taylor,$^a$ L. Balicas,$^c$ \\N.D. Zhigadlo,$^d$
S.M. Kazakov,$^d$ J. Karpinski,$^d$ J.P.H. Charmant$^e$ and J. Kortus$^f$}

\affiliation{$^a$H. H. Wills Physics Laboratory, University of Bristol, Tyndall Avenue, BS8 1TL, United Kingdom.}

\affiliation{$^b$Department of Physics and IRC in Superconductivity, University of Cambridge, Madingley Road, Cambridge
CB3 0HE, United Kingdom.}

\affiliation{$^c$National High Magnetic Field Laboratory, Florida State University, Tallahassee, Florida 32306, U.S.A.}

\affiliation{$^d$Laboratorium f\"{u}r Festk\"{o}rperphysik, ETH Z\"{u}rich, CH-8093 Z\"{u}rich, Switzerland.}

\affiliation{$^e$School of Chemistry, University of Bristol, Bristol BS8 1TS, United Kingdom.}

\affiliation{$^f$Institut de Physique et Chemie de Mat\'{e}riaux de Strasbourg, Strasbourg, France.}

\date{\today}
\begin{abstract}
We report a de Haas-van Alphen (dHvA) study of the electronic structure of Al doped crystals of MgB$_2$.  We have
measured crystals with $\sim 7.5$\% Al which have a $T_c$ of 33.6~K, ($\sim 14$\% lower than pure MgB$_2$). dHvA
frequencies for the $\sigma$ tube orbits in the doped samples are lower than in pure MgB$_2$, implying  a $16\pm2\%$
reduction in the number of holes in this sheet of Fermi surface. The mass of the quasiparticles on the larger $\sigma$
orbit is lighter than the pure case indicating a
 reduction in electron-phonon coupling constant $\lambda$. These observations are compared with band structure
 calculations, and found to be in excellent agreement.
 \end{abstract}

\maketitle

The binary compound MgB$_2$ is an unusual superconductor exhibiting a very high transition temperature ($T_c\simeq$
 39~K) and two distinct superconducting gaps. This behavior is believed to originate from a phonon mediated mechanism with different coupling strengths
to the electronic $\sigma$ and $\pi$ bands \cite{LiuMK01,ChoiRSCL02b}. The two-gap structure is only observable
experimentally because of very weak scattering between the $\sigma$ and $\pi$ bands  \cite{MazinAJDKGKV02}.

Atomic substitutions change the properties of MgB$_2$ both by increasing the scattering rates and by changing the
electron density. This provides a way to engineer its superconducting properties and also allows us to test theoretical
understanding of this material. The two elements which substitute most readily in MgB$_2$ are Al and C which replace Mg
and B respectively giving the general formula Mg$_{1-x}$Al$_x$(B$_{1-y}$C$_y$)$_2$. Both dopants add electrons to the
material, and cause $T_c$ to decrease in a similar way \cite{KortusDKG05} (i.e., $\partial T_c/\partial x \simeq
\frac{1}{2}\partial T_c/\partial y)$.

In principle, this reduction could result from two different effects. It might be expected that doping will increase
the interband scattering. Theory predicts  that as the interband scattering is increased the large gap would decrease,
while the small gap increases until they reach a common value and merge together. However, experimental studies (for
example point contact spectroscopy \cite{GonnelliDUCDSKJK05}) have shown that the two gaps remain distinct, at least
for low doping, and while doping appears to reduce the size of the large gap, the small gap remains constant.
Alternatively, the reduction of $T_c$ could follow from the effects of electron doping on the electronic structure. A
recent theoretical study \cite{KortusDKG05} has concluded that the reduction of $T_c$ with increasing Al/C substitution
can be explained by a reduction of the electron-phonon coupling constant which follows from a reduction in the density
of states produced by electron doping.

Although both Al and C doping produce a similar decrease in $T_c$, their effect on the upper critical field is very
different. Al doping causes only  a small change in $H_{c2}$ and its anisotropy, however, C doping increases $H_{c2}$
dramatically.\cite{KazakovPRMZJSBK05} Similarly, C doping increases the residual resistivity at a much higher rate than
Al doping.\cite{KazakovPRMZJSBK05,KarpinskiZSKBRPMWGDUS05} The $\sigma$ band has most weight on the B plane, whereas
the $\pi$ band has a large weight on both the B and Mg planes. It is therefore to be expected that replacing B with C
should strongly increase the scattering rates on both the $\sigma$ and $\pi$ bands, but replacing Mg with Al mainly
affects the $\pi$ band.

The de Haas-van Alphen (dHvA) effect is a powerful probe of the electronic structure of metals as it gives quantitative
\textbf{k}-resolved information on the Fermi surface properties. In this paper, we present a dHvA effect study which
quantifies the effect of the dopant on the electronic structure of Al doped MgB$_2$.

\begin{figure}
\includegraphics[width=7.0cm,keepaspectratio=true]{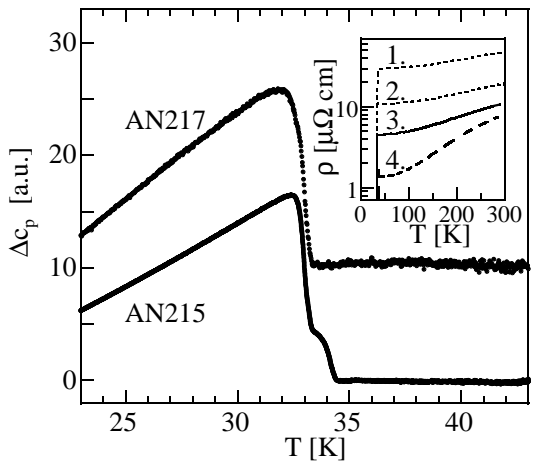}
\caption{Heat capacity vs temperature for Al doped MgB$_2$ samples AN215 and AN217. The data for AN217 have been
shifted vertically for clarity. Inset: In-plane resistivity  vs temperature, 1=C doped (AN314), 2=C doped (AN284), 3=Al
doped (AN217), 4=pure (AN77).} \label{figcp}
\end{figure}

Single crystal samples of Al and C doped MgB$_2$ were prepared in Z\"{u}rich using high pressure synthesis.
\cite{KarpinskiZSKBRPMWGDUS05} Pure MgB$_2$ crystals produced by this method have a $T_c$ of 38.5~K, which is slightly
($\sim$0.5~K) lower than the best polycrystalline samples.\cite{KarpinskiKJAPWB03} We measured two Al doped samples
from batches AN215 and AN217 which both have $T_c\simeq 33.6$~K (see below). Clear dHvA signals were observed from both
the Al doped samples for fields greater than 19~T. The Al content of the crystals was determined by measuring the
$c$-axis lattice constants which have been shown\cite{KarpinskiZSKBRPMWGDUS05} to decrease linearly with increasing $x$
($c=3.513(2)-0.28(2)x$ ~\AA). The two crystals were measured to have $c$=3.4904(7)\AA~and $c$=3.4920(10)\AA, giving
$x=7.9\pm0.4$\% and $x=7.4\pm0.4$\% for AN215 and AN217 respectively. Two C doped samples from batches AN314 and AN284
 which nominally have $y=3\%$ and $y=4\%$ with $T_c$ values of 35.7~K and 34.5~K respectively were also measured. No
dHvA signals were observed from either sample for fields up to 33~T. This is consistent with a stronger increase in
scattering rates in C doped samples, as indicated by their high residual resistances (see Fig.\ \ref{figcp} and Refs.\
\onlinecite{KarpinskiZSKBRPMWGDUS05} and~ \onlinecite{KazakovPRMZJSBK05}).

Following our dHvA study we measured the heat capacity of the two Al doped samples to probe for any homogeneous second
superconducting phases. Heat capacity is a bulk probe and hence is a useful complementary probe to the dHvA
measurements as dHvA signals will normally only be observed from the homogeneous bulk of the sample. The heat capacity
$c_p$ was measured as a function of temperature by an a.c.\ technique \cite{CarringtonMBCBMH97} in fields of up to 7~T.
The sample was placed on a flattened 12$\mu$m, chromel-constantan thermocouple and heated with a modulated light
source.

In Fig.\ \ref{figcp} we show $\Delta c_p=c_p(B=0)-c_p(B^{\|c}$=6~T) for both samples.  As the samples are in the normal
state for $B^{\|c}\gtrsim 3.5$~T, this procedure isolates that part of $c_p$ arising from the superconducting
transition. The sample labelled AN217 was the actual sample (mass=1.4$\mu$g) used for the dHvA study, whereas sample
labelled AN215 (mass=70$\mu$g) was part of the same crystal from which the much smaller dHvA sample was broken off.
Sample AN217 has a sharp single transition, with $T_c$=33.6 K (onset), (10-90\% width = 0.9~K), and
$dT_c/dB^{\|c}=-8.5\pm0.1$ K/T. Sample AN215 has more structure in its transition with a shoulder corresponding to
approximately 25\% of the total. $T_c$=34.4 K, (onset) (10-90\% width = 1.45~K) and $dT_c/dB^{\|c}=-8.8\pm0.1$ K/T,
although the onset of the main part of the transition is at approximately the same temperature as AN217. Measurements
on a pure MgB$_2$ sample (batch AN189) gave $T_c$=38.3 K, (onset) (10-90\% width = 0.3~K) and
$dT_c/dB^{\|c}=-9.8\pm0.1$ K/T. The smaller $dT_c/dB$ values in the Al doped samples almost exactly scale with their
reduced $T_c$.

Although our a.c.\ technique is very sensitive even for very small samples, it does not give accurate absolute values
of $c_p$ and further for the smallest samples there is a large addenda contribution. For these reasons the units of
$c_p$ are not quoted. For sample AN215 the jump in $c_p$ at $T_c$ is 15\% of the total which is the same as that found
for the pure MgB$_2$ samples. For AN217 the jump is $\sim 3$\%, which results from the much larger addenda relative to
the contribution from this tiny sample, rather than a reduced superconducting volume fraction.

Quantum oscillations were measured using piezoresistive cantilevers\cite{CooperCMYHBTLKK03} connected to a room
temperature a.c.\ Wheatstone bridge.  The torque is related to the change in bridge resistance by a factor $\approx
10^{-10}$Nm$\Omega^{-1}$, with a noise level roughly corresponding to a torque of $\sim 10^{-14}$Nm.

\begin{figure}
\includegraphics[width=6.0cm,keepaspectratio=true]{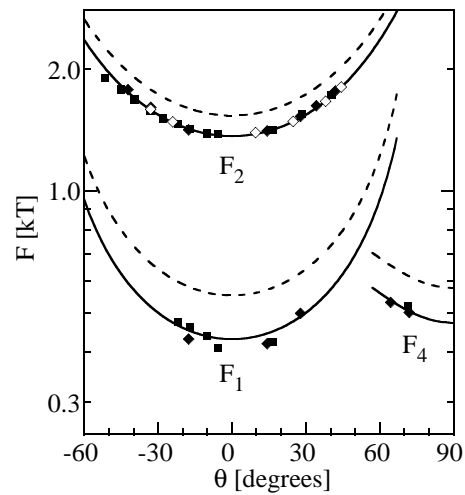}
\caption{Observed dHvA frequencies versus angle for the two Al doped samples AN215 ($\blacksquare$,$\blacklozenge$) and AN217
($\lozenge$). Dashed lines are $F(\theta)$ for pure MgB$_2$, solid lines are pure $F(\theta)$ data scaled to fit the current
data.} \label{figrot}
\end{figure}

The cantilever resistance was measured as the field was swept  (typically at 0.8~T/min) between 20~T and 33~T. Sweeps
were made at fixed angle $\theta$ as the sample was rotated from $B\|c$ to (approximately) $B\|a$.  The raw data were
fast Fourier transformed to evaluate the frequencies $F$ of any dHvA oscillations present.  In Fig.\ \ref{figrot} we
show the observed dHvA frequencies versus angle for the two Al doped samples. Comparing these results to those from
pure MgB$_2$,\cite{YellandCCHMLYT02,CarringtonMCBHYLYTKK03} we find that the two higher amplitude signals originate
from the minimal and maximal extremal orbits of the smaller of the two quasi-two-dimensional $\sigma$ sheets of the
Fermi surface\cite{KortusMBAB01}, with frequencies labelled previously $F_1$ and $F_2$ respectively
\cite{YellandCCHMLYT02,CarringtonMCBHYLYTKK03}. The values of $F_1(\theta=0)$ and $F_2(\theta=0)$ are the same in both
Al doped crystals, and are $\sim 10-20\%$ smaller than the corresponding values for pure MgB$_2$ (see Table
\ref{freqtable}). Studies on many different samples of pure MgB$_2$ (from different sources) have remarkably
reproducible dHvA frequencies (typically to within 1\%) and so the smaller values found here are significant. The
$\sigma$ sheets have an approximately circular in-plane cross-section and a cosine $c$-axis dispersion and so we can
calculate their volumes from the measured extremal areas. We therefore estimate that there are $16\pm2$\% less holes in
our Al doped samples compared to pure MgB$_2$.

Calculations\cite{Harima02,RosnerAPD02,MazinK02} of the dHvA frequencies in pure MgB$_2$  are slightly different from
the experimental values. It was shown\cite{RosnerAPD02,MazinK02,CarringtonMCBHYLYTKK03} that all the frequencies can be
brought into agreement with theory by rigidly shifting  both $\sigma$ bands down by $\sim$80~meV and both $\pi$ bands
up by $\sim$60~meV relative to the Fermi level. For our Al doped samples we find that both $\sigma$ band frequencies
can be explained by a $\sim$150~meV shift downward relative to the calculations for the pure material (or a
$\sim$70~meV shift relative to the experimental results for the pure samples).

To describe the effect of doping theoretically we have used the the virtual crystal approximation (VCA) to calculate
the electronic band structure self-consistently on a very dense $47\times47\times35$ \textbf{k}-mesh.  The dHvA
frequencies were then calculated as described previously. \cite{MazinK02}  Here the effect of doping with Al is
simulated by replacing the Mg atom with a virtual atom with charge $Z=xZ_{\rm AL}-(1-x)Z_{\rm Mg}$. In order to account
for the difference between the dHvA areas for pure MgB$_2$ and the calculations, we applied the same constant shift for
pure MgB$_2$ as described above to all doping concentrations. The VCA calculations give the following dependence of
dHvA frequencies on doping $x$, $F_1=540-1820x$, $F_2=1530-2050x$. Hence, the observed frequencies ($F_1$ and $F_2$)
correspond to a doping of $x=7.5\pm 1$\% and $8.4\pm1$\% respectively. These values compare favorably to the $x$ values
deduced from the $c$-axis lattice constant ($\overline{x}=7.7\pm0.4$\%). Close to $x=0$ the VCA calculations (including
phonon renormalization) \cite{KortusDKG05} predict $dT_c/dx=-0.50$ K/\%, and hence for our doping we expect a $T_c$
reduction of $4.0\pm0.3$~K. The actual $T_c$ reduction was 3.9~K, and so we conclude that the VCA results accurately
describe the effect of (relatively light) doping in MgB$_2$.

In sample AN215, for angles close to $\theta=90^\circ$, a further frequency ($F_4$) is observed which originates from
the hole-like $\pi$ sheet. This frequency is also smaller than that of pure MgB$_2$, however, the relatively large
error precludes any estimate of the doping on this sheet.  We note that no signals were observed from the electron-like
$\pi$ sheet, which in pure MgB$_2$ produces the strongest dHvA signal $F_3$ (Refs.\
\onlinecite{YellandCCHMLYT02,CarringtonMCBHYLYTKK03}).

\begin{table}
\begin{ruledtabular}
\caption{Calculated (LDA)\cite{MazinK02} and measured dHvA frequencies [$F^0\equiv F(\theta=0)$] for
pure\cite{CarringtonMCBHYLYTKK03} and Al doped MgB$_2$. $\Delta E$ is the rigid band shift needed to bring the theoretical values
in line with experiment.} \label{freqtable}
\begin{tabular}{cccccccc}
    &  LDA&  \multicolumn{2}{c}{Pure} & \multicolumn{2}{c}{AN215}  & AN217 \\
Orbit    &  $F^0$ &  $F^0$ &$ \Delta E$ &  $F^0$ &  $\Delta E$ &$F^0$ \\
    &[T]&[T]&[meV]&[T]&[meV]&[T]\\
 \hline
$F_1$ &   730 &  546$\pm$20 & 85 &  410$\pm$20 &  148     &$\cdots$ \\
$F_2$ &  1756 & 1533$\pm$20 & 93 & 1360$\pm$20 &  147     & 1360$\pm$20 \\
$F_3$ &  2889 & 2685$\pm$20 &-75 & $\cdots$    & $\cdots$ &$\cdots$\\
$F_4$ &   458 &  553$\pm$10 &-45 &  480$\pm$40 &-10       & $\cdots$ \\
\end{tabular}
\end{ruledtabular}
\end{table}

\begin{table}
\begin{ruledtabular}
\caption{Summary of $m^*$ and $\lambda$ values for $F_2$ for several pure crystals (B and K from Ref.
\onlinecite{CarringtonMCBHYLYTKK03}), AN77 (this work), and the two Al doped samples. Note that sample K contained
isotopically pure $^{10}$B, whereas the others contain natural (mixed isotope) B.} \label{masstable}
\begin{tabular}{lccc}
Sample&$m^*$&$m_b$&$\lambda$\\\hline
B&0.65$\pm$0.01&0.305&1.12$\pm$0.03\\
K&0.61$\pm$0.01&0.305&1.00$\pm$0.03\\
AN77&0.61$\pm$0.02&0.305&0.99$\pm$0.06\\\hline
AN215&0.57$\pm$0.01&0.297&0.92$\pm$0.03\\
AN217&0.59$\pm$0.03&0.297&0.97$\pm$0.1\\
\end{tabular}
\end{ruledtabular}
\end{table}

The quasiparticle effective mass $m^*$  on orbit $F_2$ was determined by measuring the temperature dependence of the
amplitude of the dHvA oscillations and fitting to the usual Lifshitz-Kosevich formula.\cite{Shoenberg} These
measurements were made at $\theta=10^\circ\pm1^\circ$ and $\theta=20^\circ\pm1^\circ$ for AN215 and AN217 respectively.
A reduction of 1.5\% and 7.0\% respectively was applied to give the mass at $\theta=0$. Here we have used the fact that
for this orbit $m^*$ scales accurately with the dHvA frequency. \cite{massfreqref} Results for the two Al doped samples
along with several undoped samples are shown in Table \ref{masstable}. The amplitude of the signals from $F_1$ and
$F_4$ in these samples are too small to perform an accurate mass determination.

As we do not expect there to be strong electron-electron interactions in MgB$_2$ we can use our value of $m^*$ along
with the calculated bare band mass $m_b$ to estimate the strength of the electron-phonon coupling constant $\lambda$ on
this orbit, $\lambda=m^*/m_b-1$. Previously we have shown that for pure MgB$_2$ the values of $\lambda$ calculated by
this method are in good agreement with results of band structure calculations.
\cite{YellandCCHMLYT02,CooperCMYHBTLKK03,CarringtonMCBHYLYTKK03} A complication here is that $m_b$ varies slightly with
the doping.  In Table \ref{masstable} we show the VCA calculated band masses for the pure and Al doped samples, along
with the calculated $\lambda$ values. Although there is some variation between the values for the pure samples,
$\lambda$ for the AN215 is $10\pm5$\% smaller (the error for AN217 is too large for  meaningful comparison).

The theoretical electron-phonon mass enhancement on the $\sigma$ bands is the sum of intraband and interband
  parts, which for the undoped case are $\lambda_{\sigma\sigma}=1.02$
 and $\lambda_{\sigma\pi}=0.21$ respectively, giving $\lambda_{\rm tot}^\sigma=1.23$ (note that $\lambda_{\rm tot}^\sigma$ does not
 vary substantially within each $\sigma$ sheet and only varies by $\sim7\%$ between the two  $\sigma$ sheets \cite{MazinK02}).
 For $x$=7.7\% the VCA calculation gives $\lambda_{\rm tot}^\sigma=1.16$, which is $\sim 5\%$ smaller than for $x=0$.
 This is in line with our  observations.

\begin{figure}
\includegraphics[width=8.0cm,keepaspectratio=true]{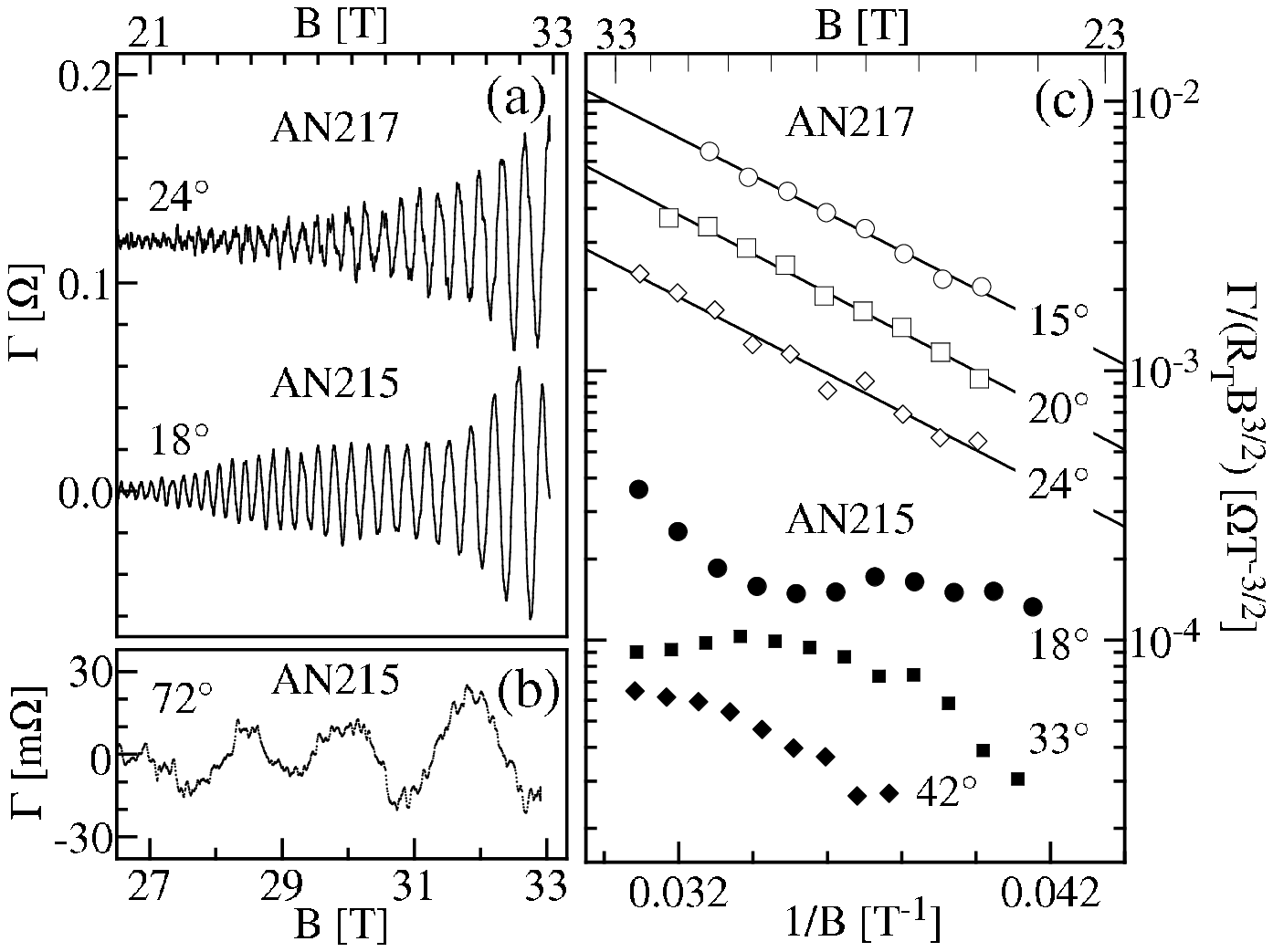}
\caption{(a) Field dependence of the oscillatory torque (Fourier filtered to remove frequencies below 1kT). AN215 data have been
multiplied by 3 for clarity. (b) Raw oscillatory torque for fields close to the basal plane showing the contribution from $F_4$
(c) Dingle plots for $F_2$ for both samples.} \label{figdingle}
\end{figure}

The field dependence of the dHvA amplitude $\Gamma_{osc}$ is proportional to $B^\frac{3}{2}R_TR_D$. Evaluating the
Dingle factor $R_D=\exp[-\pi \hbar k_F/( eB\ell)]$, where $k_F$ is the orbitally averaged Fermi wavevector, allows us
to estimate the mean-free-path $\ell$ on a particular orbit.  In Fig.\ \ref{figdingle} we show Dingle plots
[$\log(\Gamma_{osc}/(B^\frac{3}{2}R_T)$ versus $1/B$] for the frequency $F_2$ in both samples at selected angles. The
data for AN217 is exponential and is fitted by $R_D=\exp(-165/B)$, giving $\ell=270$~\AA.  This is approximately half
of that found for pure samples of MgB$_2$ (grown with natural mixed-isotope
boron).\cite{YellandCCHMLYT02,CarringtonMCBHYLYTKK03} The data for AN215 is markedly non-exponential and is consistent
with there being a beat between two different dHvA signals with frequencies differing by $\sim$40~T. As such behavior
was not found for AN217 or pure MgB$_2$ this indicates that there are two macroscopic parts to AN215 which have
different Al contents, in agreement with the specific heat measurements.\cite{40Tnote} If two frequencies are included
in the fit to the data then the scattering rate is found to be close to that of AN217.

The small amplitude of $F_4$ prevents us from determining $\ell$ accurately, however its relative amplitude suggests an
approximate two-fold decrease in $\ell$.  The signal from the hole-like $\pi$ sheet (frequency $F_3$) is normally large
for fields oriented near the plane,\cite{YellandCCHMLYT02,CarringtonMCBHYLYTKK03} and as we have not observed this
orbit in either Al doped sample we conclude that that the mean free path on this sheet must be reduced by a somewhat
larger factor ($\gtrsim 3$).

In conclusion, we have measured dHvA oscillations in Al doped MgB$_2$. Our results shows that the reduction in size of
the $\sigma$ sheets is in good agreement with band structure calculations using the VCA approximation.  The measured
reduction in quasiparticle effective mass is also consistent with the reductions in band mass and electron phonon
coupling predicted by the VCA calculations. These calculations also correctly explain the magnitude of the reduction of
$T_c$ found in our samples. As expected Al doping increases the scattering more on the $\pi$ bands than on the $\sigma$
bands. This excellent agreement between the dHvA results and theory can be seen as a direct confirmation of the effect
of band filling on the superconducting properties which has been proposed theoretically.  \cite{KortusDKG05}

We thank I.I.\ Mazin for useful discussions and B.J.\ Pullum and G.M.\ Armstrong for technical assistance. This work
was supported by the EPSRC (U.K.), and the NSF through grant number NSF-DMR-0084173. LB acknowledges support from the
NHMFL in-house program.


\begin{thebibliography}{19}
\bibitem[{40T()}]{40Tnote}
\bibinfo{note}{Note that by scaling $F_2$ with the decrease in the bulk $T_c$,
  a 40~T difference in $F_2$ corresponds to a difference in $T_c$ of $\sim$1~K
  in agreement with the $c_p$ data.}
\end{thebibliography}
\end{document}